\documentclass[twocolumn,aps,prd,superscriptaddress,longbibliography]{revtex4-2}
\usepackage{graphicx}
\usepackage{color}
\usepackage{amsmath,mathtools}
\usepackage{enumitem}
\usepackage{amssymb}
\usepackage{hyperref}
\usepackage[normalem]{ulem}
\usepackage{adjustbox}
\usepackage{nicefrac,physics}
\usepackage{stackengine}
\usepackage{lipsum}
\usepackage{multirow}
\usepackage{array}
\usepackage{feynmp-auto}
\usepackage{comment}

\newcommand{\sgn}{\rm{sgn}}

\usepackage{cancel}
\hypersetup{
	colorlinks = true,
	linkcolor = blue,
	citecolor = blue,
	urlcolor  = blue,
}

\renewcommand{\i}{{\imath}}

\begin{document}
\title{Lyapunov exponents in a Sachdev-Ye-Kitaev-type model with population imbalance in the conformal limit and beyond}

	\author{A. S. Shankar}\email[]{shankar@lorentz.leidenuniv.nl}
	\affiliation{Instituut-Lorentz, Universiteit Leiden, P.O. Box 9506, 2300 RA Leiden, The Netherlands}
	\author{M. Fremling}
	\affiliation{Institute for Theoretical Physics and Center for Extreme Matter and Emergent Phenomena,
		Utrecht University, Princetonplein 5, 3584 CC Utrecht, The Netherlands}
  	\author{S. Plugge}
   	\affiliation{Instituut-Lorentz, Universiteit Leiden, P.O. Box 9506, 2300 RA Leiden, The Netherlands}
	\author{L. Fritz}\email[]{l.fritz@uu.nl}
	\affiliation{Institute for Theoretical Physics and Center for Extreme Matter and Emergent Phenomena,
		Utrecht University, Princetonplein 5, 3584 CC Utrecht, The Netherlands}

\begin{abstract}
The Sachdev-Ye-Kitaev (SYK) model shows chaotic behavior with a maximal Lyapunov exponent. In this paper, we investigate the four-point function of a SYK-type model numerically, which gives us access to its Lyapunov exponent. The model consists of two sets of Majorana fermions, called A and B, and the interactions are restricted to being exclusively pairwise between the two sets, not within the sets. We find that the Lyapunov exponent is still maximal at strong coupling. Furthermore, we show that even though the conformal dimensions of the A and B fermions change with the population ratio, the Lyapunov exponent remains constant, not just in the conformal limit where it is maximal, but also in the intermediate and weak coupling regimes. 
\end{abstract}
\maketitle

\section{Introduction}
Over the last decade, the Sachdev-Ye-Kitaev (SYK) model has been established as a paradigmatic model accounting for a variety of phenomena ranging from aspects of the physics of black holes to non-Fermi liquids~\cite{Chowdhury-RMP2022,Rosenhaus2019-review,Franz2018-review,patel_quantum_2017,tikhanovskaya2022maximal}.
There exist two main variants of this model in the literature: one that is formulated in terms of $N$ 'complex' Dirac fermions,
and another one written in terms of $N$ 'real' Majorana fermions.
In both cases, the fermions interact via random four-body terms.
Irrespective of the formulation, one of the main features of the model is that it exhibits emergent conformal symmetry in the infrared in the strong-coupling and large-$N$ limit.
The scaling dimension of the fermion correlation function is given by $\Delta = \frac{1}{4}$ \cite{maldacena_comments_2016,polchinski_spectrum_2016},
indicative of strong interactions (for comparison, a free fermion has scaling dimension $1/2$). 

There has been a variety of proposals for the creation of SYK-like models in laboratory setups.
They range from mesoscopic systems hosting Majorana modes~\cite{Pikulin2017,Chew2017},
or Dirac fermions in graphene flakes~\cite{Chen2018,can_charge_2019},
to ultracold atomic systems \cite{danshita2017creating,wei2021optical}.
A comprehensive review of such possible setups can be found in Refs.~\cite{Chowdhury-RMP2022,Franz2018-review} and references therein.

\begin{figure}[h]
    \centering
    \includegraphics[width=1.0\columnwidth]{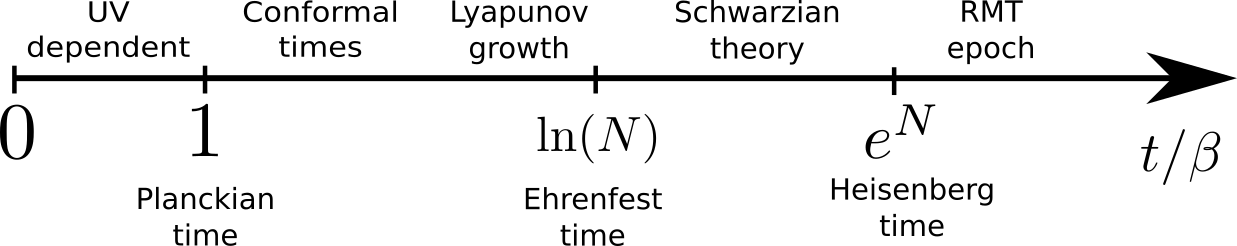}
    \caption{The SYK model exhibits multiple characteristic timescales, and with that associated regimes of dynamics.
Crucial quantities in distinguishing the different limits are the number of fermions $N$ and coupling strength $\beta J$.
This paper studies the region characterized by Lyapunov growth.
    }
    \label{fig:TimeLine}
\end{figure}

The SYK model involves three important time scales, as shown in Fig.~\ref{fig:TimeLine} (henceforth,
we measure time $t$ in units of $\beta$ and set $\hbar=k_b=1$).
They are called the Planckian time~\cite{Zaanen2019,hartnoll2021planckian,patel_magnetotransport_2018,hartnoll2022colloquium}, $t_P$,
the Ehrenfest time~\cite{gu2019relation,larkin1969quasiclassical,hashimoto_out--time-order_2017,kobrin_many-body_2021,craps_lyapunov_2020}, $t_E$,
and the Heisenberg time $t_H$. The shortest time scale, $t_P$,
is set by the condition $t_P/\beta\approx 1$. For times shorter than $t_P$,
we expect non-universal physics determined by processes at the cutoff scale.
For $t_P<t<t_E$, the dynamics is governed by the conformal mean-field theory.
The chaotic behavior associated with Lyapunov growth~\cite{stanford_many-body_2016,maldacena_bound_2016} in this regime is due to leading irrelevant operators of order $1/N$ beyond mean-field.
The Ehrenfest time is given as $t_E/\beta \approx \ln N$, where $N$ is the number of fermions.
The dynamical behavior for $t_E<t<t_H$ ceases to be described by mean-field theory plus corrections and the associated description is in terms of the Schwarzian theory of black holes.
Eventually, there is the Heisenberg time, $t_H/\beta \approx e^N$.
For times longer than $t_H$, the dynamics is described by random matrix theory. 

In this paper we study a related model, introduced in Ref.~\cite{Fremling_2022,fremling_bipartite_2021}, which emerges as a Majorana variant of the SYK model. 
It is called the bipartite SYK (or b-SYK) model and, as explained in Sec.~\ref{sec_model},
can be seen as a restricted version of the standard SYK model. Incidentally,
Majorana or complex fermion versions of similar models also appear as a natural way to incorporate internal symmetries in SYK models~\cite{lantagne2020diagnosing,Kim2019,sahoo_traversable_2020},
or to couple two or more SYK models~\cite{chowdhury_translationally_2018}.
We are interested in times shorter than the Ehrenfest time $t_E$, and mostly focus on the chaotic behavior. Furthermore, we are interested in studying the growth of the four-point function not just in the full conformal limit at strong coupling, but also at intermediate and weak couplings, as these might be relevant for experimentally achievable values of coupling and temperature, as the b-SYK model has been shown to be realizable in a laboratory by straining a real material in Ref.\cite{fremling_bipartite_2021}.

We show that the Lyapunov exponent is maximal in the conformal limit, just as for the SYK model~\cite{stanford_many-body_2016,maldacena_bound_2016,maldacena_comments_2016}. The behavior of the chaos exponent for a general number of majorana fermions in the $A$ and $B$ subsets of the b-SYK at finite coupling is unanswered in the existing literature and is the subject of the present study.
We use numerical methods to solve the Schwinger-Dyson and Bethe-Salpeter equations that are needed to extract the Green functions and Lyapunov exponents, respectively.
We find that the b-SYK model ratio of $A$ and $B$ majoranas does not influence the Lyapunov exponent for all values of coupling.

The present paper is organized as follows:
In Sec.~\ref{sec_model}, we introduce the b-SYK model and comment on how it is related to more common variants of SYK models.
In Sec.~\ref{sec_greens} we discuss the two-point functions in and away from the conformal limit. In Sec.~\ref{sec_four_point},
we compute the four-point function and introduce the equations that allow us to extract the Lyapunov exponents. In Sec.~\ref{sec_Results},
we numerically find the Lyapunov exponents and show how they depend on the population balance between $A$ and $B$ Majorana fermions.

\section{Model and methods}\label{sec_model}

\subsection{The bipartite SYK model}

The bipartite SYK (b-SYK) model consists of two sets of Majorana fermions, labelled $A$ and $B$,
with random interactions between pairs of $A$ and pairs of $B$ fermions.
Interactions between only $A$ or only $B$ fermions are absent,
and the fermion parity in both the $A$ and $B$ subsets is conserved.
The Hamiltonian reads
\begin{equation}
    H = \frac{1}{4}\sum_{ij,\alpha\beta}J_{ij\alpha\beta}\gamma^A_i\gamma^A_j \gamma^B_\alpha \gamma^B_\beta~.
\end{equation}
To distinguish the two sets of fermions we use latin indices $i,j$ for the $A$-flavor Majorana fermions ($\gamma_i^A$),
and greek indices $\alpha,\beta$ for $B$-flavor Majorana fermions ($\gamma_\alpha^B$).

We allow for $N_A$ Majorana fermions of the $A$-type and $N_B$ of the $B$-type.
The ratio $\kappa=N_A/N_B$ accounts for the relative size of the two sets.
The couplings $J_{ij\alpha\beta}$ are random and only act between sets, not within each set.
Concerning the normalization of the interaction strength, we follow the convention of Gross and Rosenhaus~\cite{gross_generalization_2017} and choose the variance of the coupling constant to be~\footnote {note that this is the $q=2, f=2$ limit of Ref.~\cite{gross_generalization_2017}}
\[
\langle J_{ij\alpha\beta} J_{i^{\prime}j^{\prime}\alpha^{\prime}\beta^{\prime}}\rangle=
\frac{J^2(N_A+N_B)}{N_A^2N_B^2}\delta_{i,i^{\prime}}\delta_{j,j^{\prime}}\delta_{\alpha,\alpha^{\prime}}\delta_{\beta,\beta^{\prime}}.
\] 

In this work, we will define $N$ as the geometric mean of $N_A$ and $N_B$, $N=\sqrt{N_{A}N_{B}}$.
We can then rewrite $\frac{N_A+N_B}{N_A^2N_B^2}=(\sqrt{\kappa} + \frac{1}{\sqrt{\kappa}})/N^3$,
which makes the symmetry between $\kappa$ and $1/\kappa$ apparent.
For clarity, this convention differs from the one used in Refs.~\cite{Fremling_2022,fremling_bipartite_2021}, where 
$
\langle J_{ij\alpha\beta} J_{i^{\prime}j^{\prime}\alpha^{\prime}\beta^{\prime}}\rangle=
\frac{J^2}{2\sqrt{N_A N_B}^3}\delta_{i,i^{\prime}}\delta_{i,i^{\prime}}\delta_{j,j^{\prime}}\delta_{\alpha,\alpha^{\prime}}\delta_{\beta,\beta^{\prime}}\;.$

The model has a well-defined large-$N$ conformal limit upon taking $N_A,N_B\to\infty$, keeping the ratio $\kappa = \frac{N_A}{N_B}$ fixed.
Rather than a single scaling dimension as in the standard SYK model, the two sets of Majorana fermions,
$A$ and $B$, have distinct scaling dimensions, $\Delta_A$ and $\Delta_B$.
These depend on the parameter $\kappa$, cf. Ref.~\cite{Fremling_2022}, as
\begin{equation} \label{eq_scalin_dims}
  \kappa = \frac{2 \Delta_A}{1-2\Delta_A}\left( \frac{1}{\tan \left( \pi \Delta_A\right)}\right)^2.
\end{equation}
For $\kappa=1$ we find $\Delta_A=\Delta_B=1/4$, just like in the standard SYK model,
although the model is still different since not all Majorana fermions interact with each other.
For other values of $\kappa$, both scaling dimensions interpolate between $0$ and $1/2$ while always fulfilling $\Delta_A+\Delta_B=1/2$.
Tunable scaling dimensions have also been found in other variants of the SYK model e.g Ref.~\cite{Marcus2019,Kim2019,garcia-garcia_sparse_2021,xu_sparse_2020}.

\subsection{Schwinger-Dyson equations}
\label{sec_greens}
\begin{figure}
    \centering
    \includegraphics[width=0.8\columnwidth]{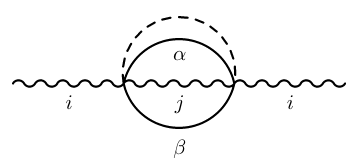}
    \includegraphics[width=0.8\columnwidth]{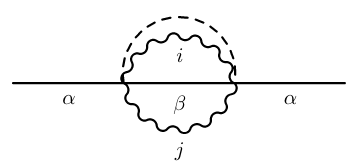}
    \caption{The diagrams that contribute to the self energies of A (top) and B (bottom) Majoranas in the large-$N$ limit. Wiggly (solid) lines denote A (B) Majorana propagators,
and the dotted line indicates a quenched disorder average $\sim J^2$.}
    \label{fig:GreensFunction}
\end{figure}

For the later numerical analysis to follow, one main input is required, the Green functions.
Hence we recapitulate the crucial steps in solving the model in the large-$N$ limit via the associated Schwinger-Dyson equations.
For more details on the procedure in the present context see e.g.~Ref.~\cite{Fremling_2022}. In this part of the paper,
the focus is more on finding a reliable numerical implementation of the Green function that allows to access the conformal limit. 
The crucial step is to consider the mean-field or large-$N$ limit.
Compared to the conventional SYK model, we have to modify the limit slightly.
We take $N_A,N_B\to\infty$ while keeping $\kappa=N_a/N_B$ fixed. As in the conventional case,
there is one order $O(1)$ diagram per species of fermions, the so-called 'melon' diagrams.
These are shown in Fig.~\ref{fig:GreensFunction}.
The diagrams contain the coupling $J^2$ to all orders and exhibit an emergent conformal symmetry in the infrared, as explained below.

\subsubsection{Imaginary time formalism}
\label{sec:imaginary_time}
The discussion of equilibrium properties of the Schwinger-Dyson (SD) equations is easiest carried out in the finite-temperature imaginary time formalism.
The inverse temperature is denoted as $\beta =1/T$ ($\hbar=k_B=1$).
For the two species, the SD equations read
\begin{equation}
    G^{A/B}(\i\omega_n) = \frac{1}{-\i\omega_n - \Sigma^{A/B}(\i\omega_n)},
    \label{eq:SDeqns}
\end{equation}
where the respective self energies are given by 
\begin{subequations}
\begin{align}
    \Sigma_A(\tau) &= \frac{J^2}2(1+\frac 1\kappa)\, G^A(\tau) \left(G^B(\tau)\right)^2, \\
    \Sigma_B (\tau)&= \frac{J^2}2(1+\kappa)\, G^B(\tau) \left( G^A(\tau)\right)^2 \;.
\end{align}
\label{eq:bSYK_Equations}
\end{subequations}
Here $\omega_n=(2n+1)\pi T$ for integer $n$ are the fermionic Matsubara frequencies, whereas $\tau$ denotes imaginary time. 
The Fourier transform  between Matsubara frequencies and imaginary time is defined according to 
\begin{subequations}
\begin{align}
    G(\i\omega_n) &= \int_0^\beta e^{\i\omega_n \tau} G(\tau) \, d\tau \;,\label{eq:time2freq}\\
    G(\tau) &= \frac{1}{\beta} \sum_{\omega_n} e^{-\i\omega_n \tau} G(\i\omega_n)\;. \label{eq:freq2time} 
\end{align}
\end{subequations}
One can show analytically that the finite temperature imaginary time Green functions are given by~\cite{Fremling_2022}
\begin{eqnarray}\label{eq_finiteTanaly}
  G^{A}(\tau)&=&a \; \sgn(\tau) \left(  \frac{\pi}{\beta \sin\left(\frac{\pi \tau}{\beta} \right)}\right)^{2\Delta_A}\;,\nonumber \\
  G^{B}(\tau)&=&b\;  \sgn(\tau) \left(  \frac{\pi}{\beta \sin\left(\frac{\pi \tau}{\beta} \right)}\right)^{2\Delta_B}\;,
\end{eqnarray}
where for a given $\kappa$, the scaling dimensions $\Delta_A$ and $\Delta_B$ are related according to Eq.~\eqref{eq_scalin_dims}.

As far as the overall constants $a$ and $b$ are concerned, it is found that only the product $ab$ is uniquely determined, and not the numbers $a$ and $b$ themselves. When we assume that the self energy dominates over the free propagator, we can use the conformal ansatz in equations Eq.~\eqref{eq:bSYK_Equations} and ~\eqref{eq:SDeqns} for each of the $A$ and $B$ flavors respectively. Naively, we would expect that the two equations are sufficient to constrain the two unknowns $a$ and $b$ respectively, but it turns out the two equations are identical, and only the product is constrained. The result is 
\begin{align}
    \frac{1}{a^2 b^2} &= \frac{J^2}{2}\left(1+\frac{1}{\kappa}\right) 2\pi \frac{\cot(\pi\Delta_A)}{1-2\Delta_A} \\
    &= \frac{J^2}{2}\left(1+\kappa\right) 2\pi \frac{\cot(\pi\Delta_B)}{1-2\Delta_B} .
\end{align} 
However, in the real system, at short times, the conformal ansatz is no longer valid, and the free propagator wins over, and $G^{A/B}(\tau)$ should go as $\frac{1}{2}\sgn(\tau)$. This is sufficient to uniquely constrain the short time dynamics of the model. 

Numerically, we solve the Schwinger-Dyson equations in a self-consistent manner by repeated evaluation of the Green functions and self-energies paired with an iteration on an imaginary time grid running from $0$ to $\beta$. Eqs.~\eqref{eq:time2freq},
\eqref{eq:freq2time} and similarly for the self-energies here are recast in the form of discrete Fourier transforms,
for which there are efficient numerical algorithms such as Fast Fourier transform.
To achieve convergence, we use a weighted update of the Green functions according to $G^{new} = \frac{x}{-\i\omega_n - \Sigma} + (1-x)G^{old}$ with a small mixing parameter $x$; here $\Sigma(\i\omega_n)$ denotes the associated self-energy calculated from $G^{old}$ of the previous iteration.

\begin{figure}
  {\centering
    \includegraphics[width=1.00\columnwidth]{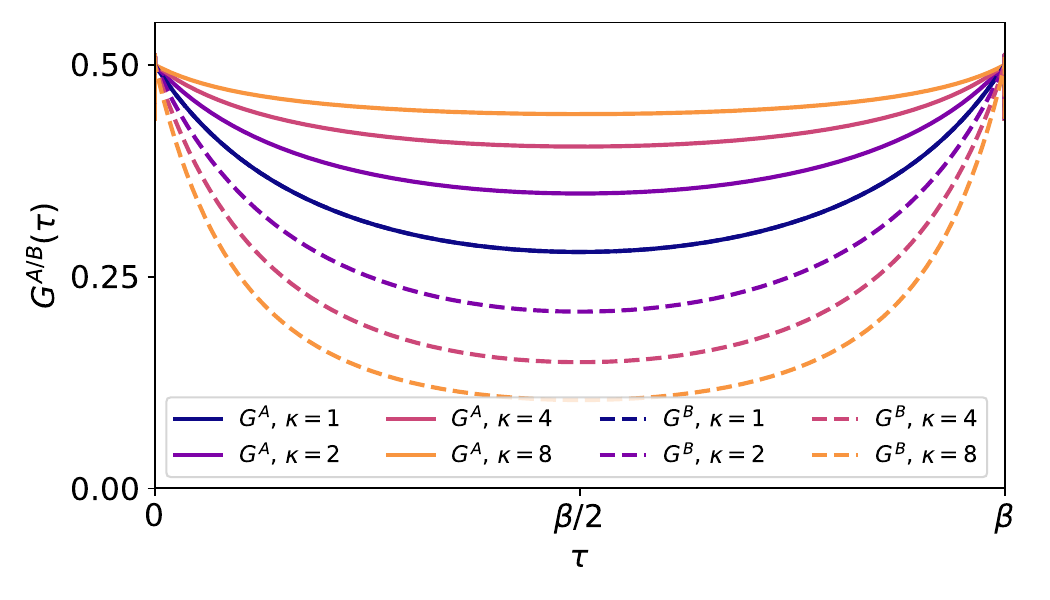} 
    \caption{Finite temperature Majorana Green functions $G^{A/B}(\tau)$ for $\beta J=10$ and several values of $\kappa$.
Taking $\kappa\to 1/\kappa$ exchanges the $A$ and $B$ species, hence we plot only $\kappa\geq1$.
    \label{fig_G_tau}}}
\end{figure}

In Fig.~\ref{fig_G_tau} we show the Majorana Green functions $G^{A/B}(\tau)$ for $\beta J=10$ and for a variety of values of $\kappa$.
By fitting the numerically obtained $G^{A/B}$ to Eq.~\eqref{eq_finiteTanaly} one can see that the scaling dimensions indeed match the conformal results. Overall,
we find excellent agreement in the region $0\ll\tau\ll\beta$.

\subsection{Real time formalism}
\label{sec:Real_Time}

The main goal of this paper is to numerically study the out-of-time-ordered correlator (OTOC) in the b-SYK model.
To compute it, we need the real time retarded Green function as input.
We first note the Dyson equation for the retarded propagator ~\cite{parcollet_non-fermi-liquid_1999,lantagne2020diagnosing,sahoo_traversable_2020,gu_notes_2020}
\begin{equation}
    \left(G^R (\omega+\i\delta)\right)^{-1} = \omega+\i\delta - \Sigma^R(\omega+\i\delta).
    \label{eq:retarded_dyson_equation}
\end{equation}
We drop the $A/B$ labels, unless explicitly required. The spectral decomposition for the Green functions reads: 
\begin{subequations}
\label{eq:SpectralDecomposition}
\begin{align}
    G(z) &= \int_{-\infty}^\infty \, \frac{d\Omega}{\pi}\frac{\rho(\Omega)}{z-\Omega},\label{eq:HilbertTransform} \\
    \rho(\omega) &= -\Im{G^R(\omega+\i\delta)}\;. \label{eq:spectraldef}
\end{align}
\end{subequations}

\begin{figure*}
  {\centering
    \includegraphics[width=0.49\linewidth]{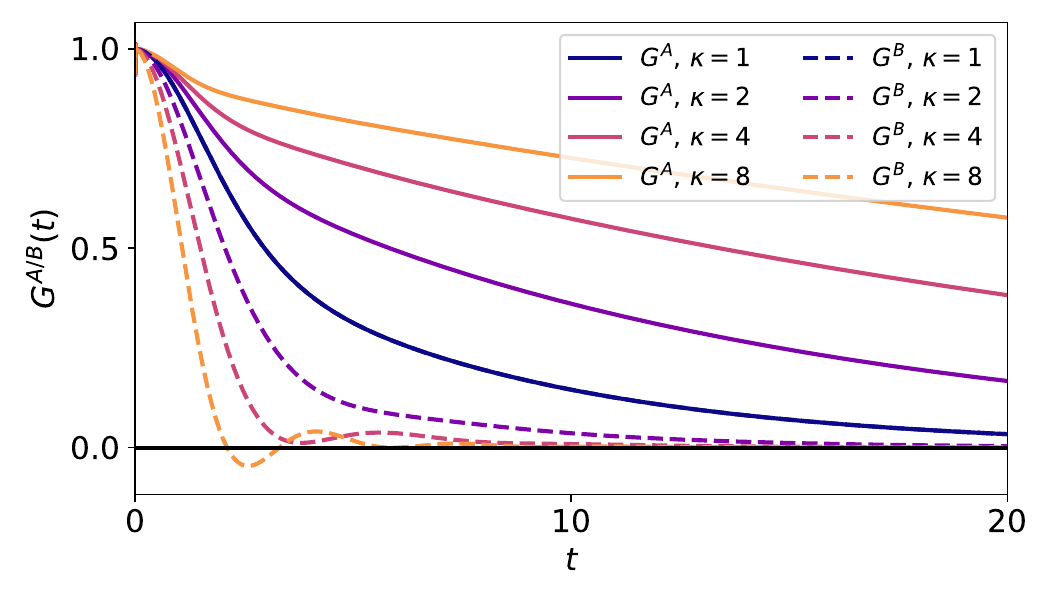}
    \includegraphics[width=0.49\linewidth]{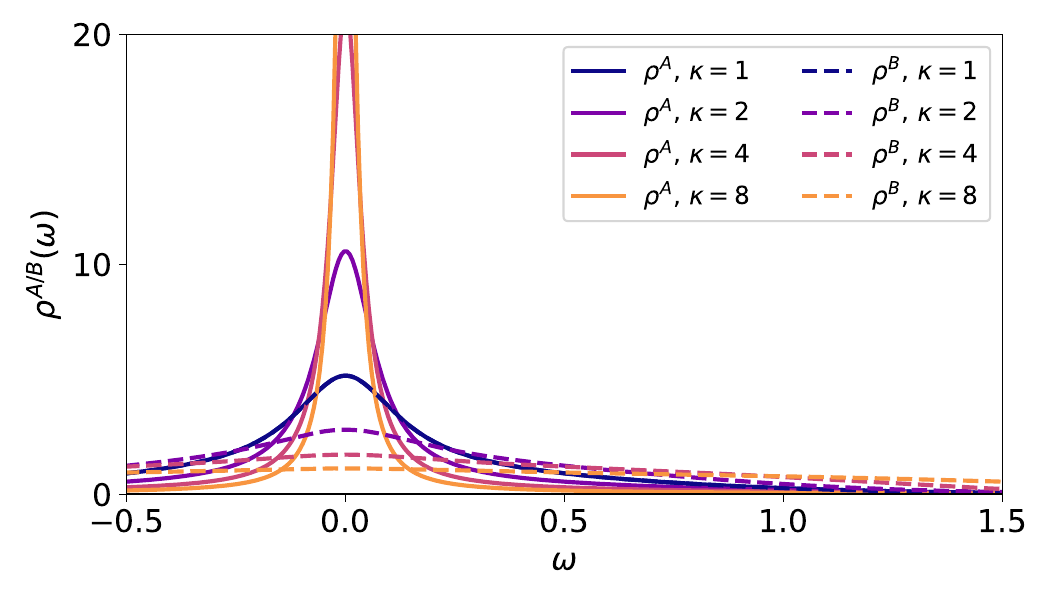} 
    \caption{
    \emph{Left panel}:
    the retarded Green functions $G_R^{A/B}(t)$ for $\beta J = 10$.
    The characteristic decay time-scale is set by the conformal dimension $\Delta_{A,B}$.
    \\\emph{Right panel}: the corresponding spectral functions, showing a strong dependence on $\kappa$.
    \label{fig_G_t}}}
\end{figure*}
Since the self energies are well defined in imaginary time according to Eq.~\eqref{eq:bSYK_Equations}, we can use Eqs.~\eqref{eq:time2freq},
\eqref{eq:freq2time} and \eqref{eq:SpectralDecomposition} to express $\Sigma(\i\omega_n)$ in terms of the spectral function.
The analytical continuation is then done by replacing $\i\omega_n \xrightarrow{} \omega+\i\delta$, resulting in
\begin{widetext}
\begin{equation}
  \Sigma^R_B(\omega+i\delta) = \frac{J^2}2(1+\kappa) \int\int\int
  \frac{\dd \omega_1}{\pi} \frac{\dd \omega_2}{\pi}\frac{\dd \omega_3}{\pi}
  \rho_A(\omega_1)\rho_A(\omega_2)\rho_B(\omega_3)
  \frac{\left[n(\omega_1)n(\omega_2)n(\omega_3) + n(-\omega_1)n(-\omega_2)n(-\omega_3)\right]}{\omega + \i\delta -\omega_1 -\omega_2-\omega_3},
\end{equation}
\end{widetext}
where $n(\omega)$ is the Fermi-Dirac distribution function.
The expression for $\Sigma_A$ is obtained by changing $A\leftrightarrow B$, and $\kappa\leftrightarrow 1/\kappa$.
In principle,
the Schwinger-Dyson equations can be solved iteratively for $G^R_{A/B}(\omega)$ and $\rho^{A/B}(\omega)$.
However, nested numerical integration is both highly inefficient in its usage of resources and numerically unstable.
Instead, it is beneficial to rewrite it using the following decomposition which allows an implementation using only the discrete Fourier transform, cf. Refs.~\cite{Plugge2020,sahoo_traversable_2020}.
We can express the self energies as
\begin{widetext}
\begin{align}
   \Sigma^R_A(\omega+\i\delta) &=  \begin{multlined}[t][] -\imath \frac{J^2}2(1+\frac1\kappa)\int_0^\infty dt\,
     e^{\i(\omega+\i\delta)t}
     \left[n^+_A(t)n^+_B(t)n^+_B(t) + n^-_A(t)n^-_B(t)n^-_B(t)\right] \end{multlined}\\
    \Sigma^R_B(\omega+\i\delta) &= \begin{multlined}[t][] -\imath \frac{J^2}2(1+\kappa)\int_0^\infty dt\,
      e^{\i(\omega+\i\delta)t}
      \left[n^+_B(t)n^+_A(t)n^+_A(t) + n^-_B(t)n^-_A(t)n^-_A(t)\right]\end{multlined}\;,
\end{align}
\end{widetext}
where the function $n^{\pm}_{A/B}(t)$ is defined through
\begin{align}
    n^{\pm}_{A/B}(t) = \int_{-\infty}^\infty \frac{d\omega_1}{\pi} e^{-\i\omega_1t}\rho_{A/B}(\omega_1)n(\pm\omega_1)\;.
\end{align}

The retarded Green function and the corresponding spectral functions obtained from the real-time/frequency iteration of the above SD equations are shown in Figure~\ref{fig_G_t}.

\section{The four-point function}\label{sec_four_point}

\begin{figure*}
  \includegraphics[width=1.0\linewidth]{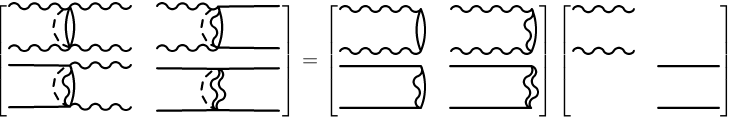}
  \caption{Diagrammatic representation of the matrix-kernel equation~\eqref{eq:kernelmatrixequation} at first order. Repeated application of the kernel $K$ generates all terms in $\mathcal F$}
  \label{fig:diagrammatic_kernel}
\end{figure*}

We now turn our attention to the four-point correlators of the b-SYK model,
and in particular to the out-of-time-ordered correlators (OTOCs).
Before we have a look into OTOCs themselves, we first discuss conventional four-point functions. In imaginary time,
a general four-point function of Majoranas has the form~\cite{gross_generalization_2017} 
\begin{equation}\label{eq:fourpoint}
    \mathcal{F}(\tau_1,\tau_2,\tau_3,\tau_4) = \frac{1}{N^2}\sum_{ijkl}\expval{\gamma^{f_1}_i(\tau_1)\gamma^{f_2}_j(\tau_2),\gamma^{f_3}_k(\tau_3)\gamma^{f_4}_l(\tau_4)}.
\end{equation}
The disorder averaging and the large-$N$ limit taken together restrict the contributions to the four-point functions to stem from what are known as ladder diagrams. These can be categorized into four channels,
depending on the flavors of the incoming and outgoing pairs of fermion propagators: AA-AA, AA-BB, BB-AA, and BB-BB.
A diagram with $n+1$ rungs can be obtained from a diagram with $n$ rungs by convolution with a kernel~\cite{stanford_many-body_2016}. In the vicinity of the Ehrenfest time $t_E$,
this can be cast as a self-consistent Bethe-Salpeter equation according to
\begin{widetext}
\begin{equation}
    \mathcal{F}_{\alpha\beta}(\tau_1,\tau_2,\tau_3,\tau_4) = \int \dd \tau \dd \tau^\prime \,K_{\alpha\gamma}(\tau_1,\tau_2,\tau,\tau^\prime)\, \mathcal{F}_{\gamma\beta}(\tau,\tau^\prime,\tau_3,\tau_4)
    \label{eq:kernelmatrixequation}
\end{equation}

where $\gamma$ is summed over, and the Kernel matrix is given as (in imaginary time and a regularized version in real time respectively)
\begin{align}
    K_{\alpha\gamma}(\tau_1\cdots \tau_4) = -J^2
    \begin{pmatrix}
    \frac{1}{2} (1+\frac{1}{\kappa})\,G^A(\tau_{13})G^A(\tau_{24})\left(G^B(\tau_{34})\right)^2   & (1+\frac{1}{\kappa})\,G^A(\tau_{13})G^A(\tau_{24})\left(G^A(\tau_{34})G^B(\tau_{34})\right) \\ 
    (1+\kappa) \,G^B(\tau_{13})G^B(\tau_{24})\left(G^A(\tau_{34})G^B(\tau_{34})\right) & \frac{1}{2} (1+\kappa)\, G^B(\tau_{13})G^B(\tau_{24})\left(G^A(\tau_{34})\right)^2
    \end{pmatrix} 
    \label{eq:kernel_tau} \\
    K^R_{\alpha\gamma}(t_1\cdots t_4) = J^2
    \begin{pmatrix}
    \frac{1}{2} (1+\frac{1}{\kappa})\,G^A_R(t_{13})G^A_R(t_{24})\left(G^B_W(t_{34})\right)^2   & (1+\frac{1}{\kappa})\,G^A_R(t_{13})G^A_R(t_{24})\left(G^A_W(t_{34})G^B_W(t_{34})\right) \\ 
    (1+\kappa) \,G^B_R(t_{13})G^B_R(t_{24})\left(G^A_W(\tau_{34})G^B_W(t_{34})\right) & \frac{1}{2} (1+\kappa)\, G^B_R(t_{13})G^B_R(t_{24})\left(G^A_W(t_{34})\right)^2
    \label{Kernelt}
    \end{pmatrix}
\end{align}
\end{widetext} 
The indices $\alpha,\beta,\gamma$ refer to the flavors of the Majorana propagators on the external legs. For example,
$F_{00}$ refers to the AA-AA scattering and $F_{10}$ refers to BB-AA scattering. 
A diagrammatic representation of the matrix-kernel equation~\eqref{eq:kernelmatrixequation} is shown in Fig.~\ref{fig:diagrammatic_kernel}.

\begin{figure}
  \centering
  \includegraphics[width=1.0\columnwidth]{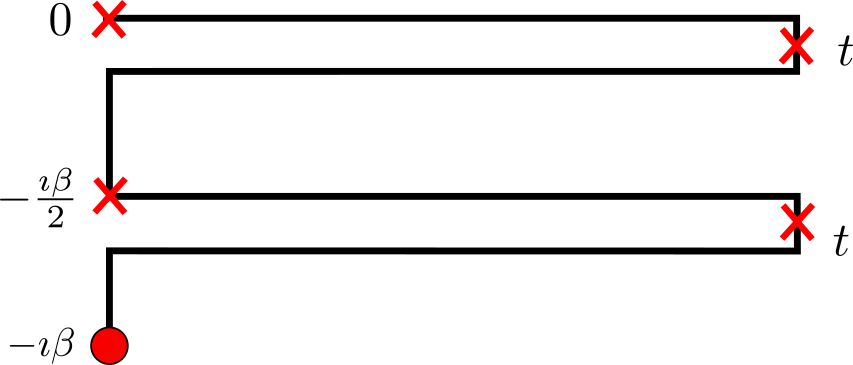}

  \caption{Schwinger-Keldysh contour with two temporal folds (excursions to time $t$) and Majorana operator insertions (red crosses) that represents the regularized OTOC in Eq.~\eqref{eq:regularized_OTOC}.}
  \label{fig:two_fold_contour}
\end{figure}

Quantum chaos is characterized by the Lyapunov exponent. Instead of looking at the real time version of Eq.~\eqref{eq:fourpoint}, we consider a regularized version according to
\begin{equation}
    F_{ab}(t_1,t_2) = \frac{1}{N^2}\sum_{a,b}\overline{\Tr{\sqrt{\rho}\comm{\gamma_a(t_1)}{\gamma_b(0)}\sqrt{\rho}\comm{\gamma_a(t_2)}{\gamma_b(0)}}}. 
    \label{eq:regularized_OTOC}
\end{equation}
This regularized OTOC has the thermal density matrix $\rho$ of the thermal average split evenly between pairs of Majorana operators, and brackets $[\cdot,\cdot]$ denote commutators.
In diagrammatic language this means that the four point function is evaluated on a double-fold Schwinger-Keldysh contour with insertions of the Majorana operators as shown in Fig.~\ref{fig:two_fold_contour}.

This is a regularization not of the UV, but of the IR. Details on which of the many possible choices of regularization and Schwinger-Keldysh contour one might pick can be found in Ref.~\cite{romero2019regularization}.
The key point is that for massless theories, which the SYK universality class belongs to,
all different regularizations give the same exponential growth, even though the values of the actual OTOCs may differ.
For the choice in Eq.~\eqref{eq:regularized_OTOC}, the four point function in question will be generated by ladder diagrams with retarded or advanced Green functions on the rails,
and so-called Wightman functions $G^W(t)=G(\frac{\beta}{2} + i t)$ on the rungs.
Formally, the latter are obtained by an analytic continuation of the imaginary time Green function noted in Sec.~\ref{sec:imaginary_time}.
This analytic continuation can be be performed with the use of the spectral decomposition, also known as a Hilbert transform.
In total, one obtains the result
\begin{equation}
    G^W(\omega) = \frac{\rho(\omega)}{2\cosh{\frac{\beta\omega}{2}}}.
\end{equation}

The late time exponential growth of the OTOC \cite{maldacena_bound_2016} can then be fit to the Lyapunov ansatz
\begin{equation}
    \mathcal{F}_{\alpha\beta}(t_1,t_2) = e^{\lambda_{\alpha\beta}\frac{(t_1 + t_2)}{2}}\, f_{\alpha\beta}(t_{12})\;.
    \label{eq:Lyapunov_Ansatz}
\end{equation}
As opposed to the standard SYK model, each of the four different scattering channels might ostensibly have its own Lyapunov exponent. It turns out that this is not the case.
A detailed technical explanation involving the consistency of the Lyapunov ansatz with a single exponent $\lambda$ is presented in Appendix~\ref{sec:technical_explanation}. 

A simple qualitative argument for a single Lyapunov exponent is that the scattering channels all feed back into each other.
The AA-AA scattering amplitude also passes through the AA-BB channel and then back into the BB-AA channel.
This imposes a sense of self-consistency between the scattering channels,
which in turn forces them to have the same late time Lyapunov growth.

\subsection{Conformal limit}
Taking the ansatz that all four Lyapunov exponents $\lambda_{\alpha\beta}$ are the same, {\it i.e.} $\lambda_{\alpha \beta}=\lambda$ allows us to make an ansatz for the growth equation. First, we will notice that the equations for $f_{00}$ and $f_{10}$ decouple, and we get the same equations for the other pair $f_{01}$ and $f_{11}$. 
In the conformal limit, following~\cite{maldacena_comments_2016} we can use the conformal mapping to obtain the retarded and Wightman Green functions from Eqs.~\eqref{eq_finiteTanaly} to get
\begin{subequations}
    \label{eq:ConfRealTimeGreens}
\begin{align}
    G_R^A(t) &= 2 a \cos(\pi \Delta_{A})\left(\frac{\pi}{\beta \sinh\frac{\pi t}{\beta}}\right)^{2\Delta_{A}} \\
    G_W^A(t) &= a \left(\frac{\pi}{\beta\cosh\frac{\pi t}{\beta}}\right)^{2\Delta_A}, 
\end{align}
\end{subequations}
and likewise for the $B-$ fermions. 
The growth ansatz can also be made in analogy with the regular SYK case: 
\begin{equation}
    \begin{pmatrix}
        f_{00}(t_{12}) \\ f_{10}(t_{12}) 
    \end{pmatrix}
    = \begin{pmatrix}
        a\,\mathcal{C}_a\left(\frac{\pi}{\beta\cosh{(t_{12}\frac{\pi}{\beta
        })}}\right)^{2\Delta_a + h} \\
        b\,\mathcal{C}_b\left(\frac{\pi}{\beta\cosh({t_{12}\frac{\pi}{\beta
        }})}\right)^{2\Delta_b + h}
    \end{pmatrix} e^{h(t_1+t_2)\frac{\pi}{\beta
        }} \label{eq:b-sykLyapunovAnsatz}
\end{equation}
It can be noted that Eq.~\eqref{eq:b-sykLyapunovAnsatz} is a way of rewriting Eq.~\eqref{eq:Lyapunov_Ansatz} in a way that is convenient for the conformal limit calculation. $\mathcal{C}_a $ and $ \mathcal{C}_b$ are hitherto undetermined constants. The equations one needs to solve are then (the factors of $\frac{\pi}{\beta}$ have been chosen appropriately so that they scale away) 
\begin{widetext}
\begin{subequations}
\begin{multline}
    e^{h(t_1 + t_2)} f_{00}(t_{12}) = \frac{J^2}{2}(1+\frac{1}{\kappa})\int dt_3 dt_4 \biggl[G^A_R(t_{13})G^A_R(t_{24})G^B_W(t_{34})^2f_{00}(t_{34}) + \\2 G^A_R(t_{13})G^A_R(t_{24})G^B_W(t_{34})G^A_W(t_{34})f_{10}(t_{34})\biggr]e^{h(t_3+t_4)}
\end{multline}
\begin{multline}
    e^{h(t_1 + t_2)} f_{10}(t_{12}) = \frac{J^2}{2}(1+\kappa)\int dt_3 dt_4 \biggl[G^B_R(t_{13})G^B_R(t_{24})G^A_W(t_{34})G^B_W(t_{34})f_{00}(t_{34}) + \\2 G^B_R(t_{13})G^B_R(t_{24})G^A_W(t_{34})^2f_{10}(t_{34})\biggr]e^{h(t_3+t_4)}
\end{multline}
\end{subequations}
\end{widetext}
The way to solve these equations is to first represent the $t_{34}$ part as an inverse fourier transform, which factorizes the integral into a function that depends only on $t_3$ and another function that depends only on $t_4$, which can be separately integrated. One can express the fourier transforms for powers of hyperbolic sines and cosines as analytic continuations of the Euler Beta function 
\begin{subequations}
    \begin{align}
    \int_{-\infty}^\infty dt\,e^{i\omega t} \frac{1}{(\cosh{t})^\alpha} &=  2^{\alpha-1} \mathrm{B}\left(\frac{\alpha - i\omega}{2},\frac{\alpha + i\omega}{2}\right) ,\\
    \int_{-\infty}^\infty dt\,e^{i\omega t} \frac{\theta(t)}{(\sinh{t})^\alpha} &=  2^{\alpha-1}\mathrm{B}\left(\frac{\alpha - i\omega}{2},1-\alpha\right) .
    \end{align}
\end{subequations}
The result then is that  
\begin{subequations}
\begin{align}
    \mathcal{C}_a &= \mathcal{M}\left(\mathcal{C}_a + 2\mathcal{C}_b\right) \\
    \mathcal{C}_b &= \mathcal{M}^\prime\left(2\mathcal{C}_a + \mathcal{C}_b\right) ,
\end{align}
\label{eq:confomalconsistency}
\end{subequations}

where 
\begin{align}
    \mathcal{M} &= \frac{(1-2\Delta_A)\sin(2\pi\Delta_A)}{\pi}\frac{(\Gamma(1-2\Delta_A))^2\Gamma(2\Delta_A + h)}{\Gamma(2-2\Delta_A + h)} \\
    \mathcal{M}^\prime &= \frac{(1-2\Delta_B)\sin(2\pi\Delta_B)}{\pi}\frac{(\Gamma(1-2\Delta_B))^2\Gamma(2\Delta_B + h)}{\Gamma(2-2\Delta_B + h)}
\end{align}
The equations Eqs.~\eqref{eq:confomalconsistency} only have a trivial solution $\mathcal{C}_A = \mathcal{C}_B = 0$ if either of the scaling dimensions are $0$ or $\frac{1}{2}$, i.e, the $\kappa = 0$ and $\kappa \rightarrow \infty$ models are not chaotic in the strictly conformal limit. 

For any other intermediate $\kappa$, even infinitesimally small, Eqs.~\eqref{eq:confomalconsistency} permit a solution if 
\begin{align}
    \det\mqty[\mathcal{M}-1 & 2\mathcal{M} \\
    2\mathcal{M}^\prime & \mathcal{M}^\prime -1] = 0. 
\end{align}
We have solved this equation for $h$ and the solution found is always $h=1$ for any value of $\kappa$. This means that for the b-SYK model, it is always possible to increase the coupling and lower the temperature sufficiently that the system always has a maximal Lyapunov exponent $\lambda = \frac{2\pi}{\beta}$. 

For realistic couplings and not too low temperatures, one needs to observe the behavior of the Lyapunov exponent including non-conformal corrections to the Green function by perturbatively including the $\i\omega$ term in the Dyson equation.  If the correction to the Kernel is $\delta K_R$, and if we compute all the eigenvalues in the conformal limit and call them $k(h)$, the we can Taylor-expand $k(h)$ about $h=1$. The point is now that $h=1$ gives eigenvalue $k(h) = 1$, so we say that 
\begin{align}
    k(1+\delta h) = 1 + k^\prime(1) \, \delta h 
\end{align}
Thus in order to keep the kernel having eigenvalue 1, the correction 
\begin{align}
    \expval{\delta K_R} &= \delta h k^\prime(1) \nonumber\\
    \implies \delta h &= \frac{\expval{\delta K_R}}{ k^\prime(1)}
\end{align}
is the first non-conformal correction to the lyapunov exponent.

\subsection{Numerical analysis for weak and intermediate coupling}

Rather than take this complicated approach, the weak and intermediate coupling limits can be analysed numerically.
We can bring the kernel equation into the concise form 
\begin{align}
    f_{\alpha\beta}(\omega) = \abs{G^\alpha_R(\omega+\i\frac{\lambda}{2})}^2\left(\tilde{K}_{\alpha 0} \ast f_{0\beta} + \tilde{K}_{\alpha 1}\ast f_{1\beta}\right)~,
    \label{eq:conv_form}
\end{align} 
where additionally a Fourier transform was performed. The ansatz function $f_{\alpha\beta}(\omega')$ is analyzed in frequency space, see below.
We also denote the shifted frequency $\tilde{\omega} = \omega + \i\frac{\lambda}{2}$ that enters in the retarded Green function.
The latter is obtained from the regular retarded Green function $G_R(\omega+\i\delta)$ that is calculated in Sec.~\ref{sec:Real_Time} by use of the Fourier shift theorem.
The symbol $\star$ in Eq.~\eqref{eq:conv_form} indicates a convolution with the ansatz function $f_{\gamma\beta}(\omega)$.
The part of the kernel elements $\tilde{K}_{\alpha\gamma}(\omega)$ that contains the Wightman Green functions is given by 
\begin{widetext}
\begin{equation}
    \tilde{K}_{\alpha\beta}(\omega) = J^2
    \begin{pmatrix}
    \frac{1}{2} (1+\frac{1}{\kappa})\,\mathfrak{F}\left[(G^B_W(t))^2\right] & (1+\frac{1}{\kappa})\,\mathfrak{F}\left[(G^B_W(t)\,G^A_W(t))\right]  \\
    (1+\kappa)\,\mathfrak{F}\left[(G^B_W(t)\,G^A_W(t))\right]  & \frac{1}{2}(1+\kappa)\,\mathfrak{F}\left[(G^A_W(t))^2\right]
    \end{pmatrix}
\end{equation}
\end{widetext}
where $\mathfrak{F}\left[ \cdot \right] $ represents the Fourier transformation.

 Finally, note that Eq.~\eqref{eq:conv_form} can be thought of as an eigenvalue problem for the ansatz $f_{\alpha\beta}(\omega)$ in frequency space $\omega$ with a block structure $\alpha,\beta$ due to the different kernel matrix blocks according to 
\begin{widetext}
\begin{equation}
\begin{bmatrix}
f_{00}(\omega)\\
f_{10}(\omega)\\
f_{01}(\omega)\\
f_{11}(\omega)
\end{bmatrix}=\begin{bmatrix}
|G_{R}^{A}(\tilde{\omega})|^{2}\tilde{K}_{00}(\omega-\omega^{\prime}) & |G_{R}^{A}(\tilde{\omega})|^{2}\tilde{K}_{01}(\omega-\omega^{\prime}) & 0 & 0 \\
|G_{R}^{B}(\tilde{\omega})|^{2}\tilde{K}_{10}(\omega-\omega^{\prime}) & |G_{R}^{B}(\tilde{\omega})|^{2}\tilde{K}_{11}(\omega-\omega^{\prime}) & 0 & 0 \\
 0 & 0 &
 |G_{R}^{A}(\tilde{\omega})|^{2}\tilde{K}_{00}(\omega-\omega^{\prime}) & |G_{R}^{A}(\tilde{\omega})|^{2}\tilde{K}_{01}(\omega-\omega^{\prime})\\
 0 & 0 &
 |G_{R}^{B}(\tilde{\omega})|^{2}\tilde{K}_{10}(\omega-\omega^{\prime}) & |G_{R}^{B}(\tilde{\omega})|^{2}\tilde{K}_{11}(\omega-\omega^{\prime})
\end{bmatrix}\begin{bmatrix}f_{00}(\omega^{\prime})\\
f_{10}(\omega^{\prime})\\
f_{01}(\omega^{\prime})\\
f_{11}(\omega^{\prime})
\end{bmatrix}\;.
\end{equation}
\end{widetext}
On the finite frequency grid, the convolution operations naturally translate to matrix multiplications.
For a solution of $f_{\alpha\beta}$ to exist, the matrix operator needs to have $1$ as its largest eigenvalue~\cite{stanford_many-body_2016,maldacena_comments_2016,gu2019relation}.
This is equivalent to saying that Eq.~\eqref{eq:Lyapunov_Ansatz} is the correct form for the late time behavior of the OTOC,
and the Lyapunov exponent is thus fixed uniquely. 
\begin{figure*}
    \includegraphics[width=1.00\columnwidth]{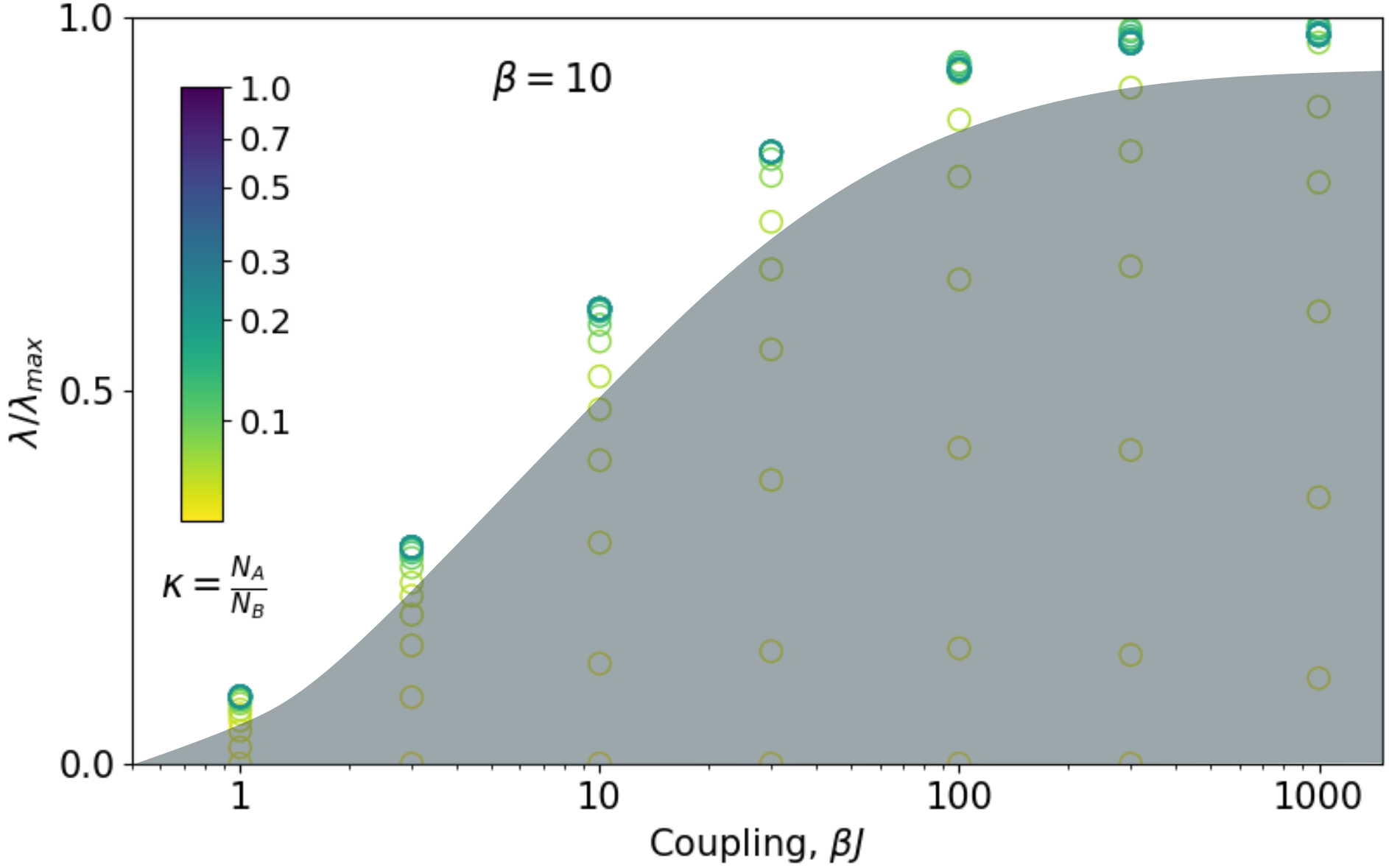}
    \includegraphics[width=1.00\columnwidth]{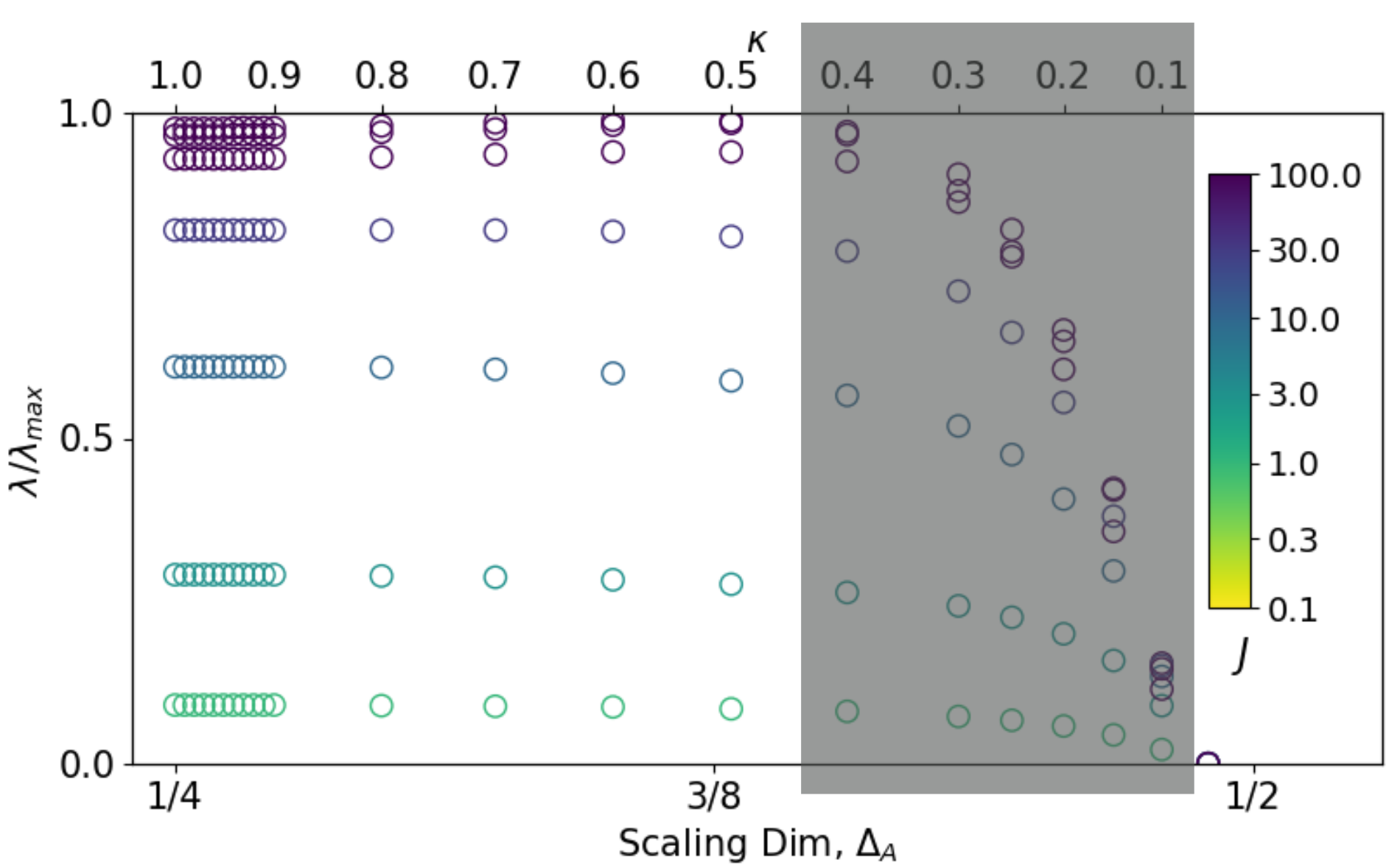}
    \caption{(Left) The Lyapunov exponent as a function of the coupling strength $\beta J$ and for various values $\kappa=N_A/N_B$.
    For $\kappa > 0.7$ and $\beta J \gtrsim 300$ the b-SYK model saturates the quantum chaos bound of $\lambda = 2\pi/\beta$.
    The special case $\kappa=1$ has identical $\lambda$ as in the SYK model.
    (Right) The Lyapunov exponent as a function $\kappa$ for various values of $\beta J$.
    We find that when $\kappa \gtrsim 0.5$ then $\lambda$ is independent of $\kappa$.
    The apparent downturn of the Lyapunov exponent, as a function of $\kappa$, can be attributed to the inability of the numerics when the scaling dimensions for the two species are drastically different.
    In both figures, the grayed out region shows where the numerical results should not be trusted. 
    }
    \label{fig_Lyaponov}
\end{figure*}

\section{Results}
\label{sec_Results}
\subsection{Analytics and numerics}
We now present and discuss the results of our numerical calculations and compare to analytically known limits. This will reveal some limitations of the numerical method rooted in numerous finite size effects. 
From the analysis in the preceding chapter, we know that the Lyapunov exponent $\lambda$ is maximal in the conformal limit for all values of $\kappa$.
Furthermore, we confirmed numerically that for $\kappa=1$, then $\lambda$, as a function of $J$, has identical behavior as in the normal SYK model.
This behavior has previously been studied in Ref.~\cite{maldacena_comments_2016}. 

Numerically, we studied the behavior of $\lambda$ as a function of $\beta J$ for various values of $\kappa=N_A/N_B$.
Figure~\ref{fig_Lyaponov} (left) shows the Lyapunov exponent $\lambda$ as a function of the coupling $\beta J$ for a variety of values of $\kappa$. The different values of $\kappa$ are encoded in the color scale. We do not show values of $\kappa>1$ because they are equivalent to those for $1/\kappa$ by symmetry upon exchange of the species.
The figure suggests that $\lambda$ for all curves with $\kappa\approx1$ are approximately the same. Smaller values of $\kappa$ seem to differ significantly in their value of $\lambda$ (the gray shaded region is affected by strong finite size effects and the results should not be trusted, see discussion in Appendix~\ref{app:FiniteSize}).
We find that the numerics allows to approach the fully conformal limit of the model, meaning $\lambda/\lambda_{\rm{max}}$ approaches $1$ in the strong coupling limit for values $\kappa \approx 1$, in agreement with our analytical results.

For intermediate couplings $\beta J$, which is beyond the reach of any analytical treatment, numerical calculations are more accurate~\cite{maldacena_comments_2016}.
Similar to Ref.~\cite{maldacena_comments_2016}, we find for this regime of $J$, that the Lyapunov exponent decreases following a $1/J$ behavior.
In total, we find that for values of $0 \ll \kappa\leq 1$, the Lyapunov exponent is mostly agnostic to the population ratio $\kappa$.

It is instructive to analyze the $\kappa$ dependence in more detail.
In Figure~\ref{fig_Lyaponov} (right) we fix $J$ and vary $\kappa$ (or $\Delta_A$).
We observe that the value of $\lambda$ is \emph{independent} of $\kappa$ up to some characteristic value of $\kappa$, after which it begins to decline (grey area).
We argue that the downturn in $\lambda$ is an artifact of the numerical method we are using. Essentially we are seeing a finite-size effect in that the time/frequency discretization in the numerics is not fine enough. We have checked for isolated points that the gray area can be pushed upon increasing the resolution. 

An immediate question that follows is why the finite-size effects appear only for values of $\kappa$ away from 1. This can be understood upon considering the scaling dimensions as a function of $\kappa$: decreasing $\kappa$ increases the spread in scaling dimensions of the $A$ and $B$ Majorana fermions.
This implies that one has to keep track  of two time/and frequency scales that we need to accurately capture with our numerical frequency-grid where the scaling limit of one of the two is pushed to larger times. Getting a good resolution of that requires a finer frequency grid at small frequencies.
When $\kappa$ deviates too much from 1 this becomes increasingly costly in terms of time/frequency steps. An extended discussion of the finite size effects in the two-fermion Green function is given in Appendix~\ref{app:FiniteSize}.

\subsection{Discussion and Conclusion}

Having established that the Lyapunov exponent is independent of $\kappa$, 
we can compare our results to a similar model presented in Ref.~\cite{chen2017tunable}.
In that case, the authors find a Lyapunov exponent in the conformal limit which can be tuned by adjusting the relative populations of the different species of fermions. In our model, we find a stark contrast to this behavior.
Instead, we find that our model's Lyapunov exponent is completely impervious to the relative number of fermion species. In the conformal limit,
aside from showing this result in an explicit analytical calculation,
we can motivate the result in a physical way, as a sort of "proof by contradiction".
If for example, the $A- $ flavor Majorana had a smaller Lyapunov exponent,
the diagrams contributing to its four point function proceed by a pathway in which they scatter into two $B- $ flavor Majoranas,
which would then propagate with the greater Lyapunov exponent,
before finally scattering back into two $A- $ flavor Majoranas. This forces both flavors to have exactly the same exponent,
and a mathematical version of this argument is presented in Appendix~\ref{sec:technical_explanation}.

The two-point function of the Majoranas are characterized by their scaling dimension,
which is quite sensitive to the relative population ratio $\kappa$,
so one would expect that the four-point function as characterized by the Lyapunov exponent would depend on $\kappa$ as well,
but we have shown conclusively that this is not the case for cases of strong, intermediate and weak coupling, which is quite surprising.
An interesting future direction of study would be to consider what deformations should be introduced to the theory in order to have a different Lyapunov exponent for the two flavors of Majoranas.

The present work on the calculation of the Lyapunov exponent in the b-SYK model shows that the features of emergent conformal symmetry and maximal quantum chaos of the SYK model are quite robust to the couplings obeying additional internal symmetries.
Besides the particular model considered here, there are many setups where parity, charge, spin,
or general flavor symmetries of the underlying fermions carry over to the interaction matrix elements~\cite{Chowdhury-RMP2022,Franz2018-review,Kim2019,sahoo_traversable_2020,xu_sparse_2020}.
The methods used here readily carry over to those models and can be applied to the calculation of Lyapunov exponents and, in general, to the analysis of Bethe-Salpeter equations.

\begin{acknowledgments}
We acknowledge discussions with Y. Cheipesh, A. Kamenev, K. Schalm, M. Haque, and S. Sachdev. Extensive discussions with D. Stanford about the conformal limit of the OTOC are also acknowledged.
SP thanks E. Lantagne-Hurtubise, O. Can, S. Sahoo, and M. Franz for many useful discussions related to SYK models and holography.  This work is part of the D-ITP consortium,
a program of the Netherlands Organisation for Scientific Research (NWO) that is funded by the Dutch Ministry of Education, Culture and Science (OCW).
SP received funding through the European Research Council (ERC) under the European Union's Horizon 2020 research and innovation program.
\end{acknowledgments}

\section*{Author Contributions}
A.S.S and M.F contributed equally to this work.

\appendix

\section{Mathematical consistency of the Lyapunov ansatz}
\label{sec:technical_explanation}

The following short consideration for the diagram piece $\mathcal{F}_{00}$ shows why we expect only one `global' Lyapunov exponent for all scattering channels.
The other components of the four-point function can be treated with exactly the same argument. The starting point is
\begin{multline}
    \mathcal{F}_{00}(t_1,t_2) = \int dt_3 dt_4\, K_{00}(t_1,t_2,t_3,t_4)\mathcal{F}_{00}(t_3,t_4) \\ + K_{10}(t_1,t_2,t_3,t_4)\mathcal{F}_{10}(t_3,t_4)
\end{multline}
where we use the definition 
\begin{align}
    t_{1,2} &= t \pm \frac{1}{2}t_{12} \nonumber \\
    t_{3,4} &= \tilde{t} \pm \frac{1}{2}t_{34} \;.
\end{align}
The factors of a half were included to keep the area element invariant under this transformation,
$dt_3dt_4 = d\tilde{t}dt_{34}$. After some algebra, for the ansatz $f_{00}$ one finds
\begin{widetext}
\begin{multline}
    f_{00}(t_{12}) = 
    J^2\frac12 \left(1+ \frac1\kappa\right) \int d\tilde{t}dt_{34} G^R_A(t_{13})G^R_A(t_{24})\Big{[}\frac{1}{\kappa}\left(G^W_B(t_{34})\right)^2 e^{\lambda_{00}\tilde{t} - \lambda_{00} t}f_{00}(t_{34}) + \\  \left(G^W_A(t_{34})G^W_B(t_{34})\right) e^{\lambda_{10}\tilde{t} - \lambda_{00} t}f_{10}(t_{34}) \Big{]}
\end{multline}
\end{widetext}
Now we Fourier transform according to 
\begin{equation}
    G^W_A(t_{34}) = \int \frac{\dd\omega_a}{2\pi} e^{-\i\omega_a t_{34}} G^W_A(\omega_a)\;.
\end{equation}
\begin{widetext}
If we calculate a sample term $f_{00}$ to illustrate the point, 
\begin{multline}
    f_{00}(\omega) = J^2\frac12 \left(1+ \frac1\kappa\right)\int \dd t_{12} e^{\i\omega t_{12}}\int \dd \tilde{t}\int \dd t_{34} \int \frac{\dd\omega_a}{2\pi} e^{-\i\omega_a(t - \tilde{t} + \frac{1}{2}(t_{12}-t_{34}))}\int \frac{\dd\omega_b}{2\pi} e^{-\i\omega_b(t - \tilde{t} - \frac{1}{2}(t_{12}-t_{34}))} \\
    G^R_A(\omega_a)G^R_A(\omega_b)\int\frac{\dd\omega_c}{2\pi}\int\frac{\dd\omega^\prime}{2\pi} e^{-\i(\omega_c +\omega^\prime)t_{34}}
    \left[\tilde{K}_{00}(\omega_c)f_{00}(\omega^\prime) e^{\lambda_{00}\tilde{t} - \lambda_{00}t} + \tilde{K}_{10}(\omega_c)f_{10}(\omega^\prime) e^{\lambda_{10}\tilde{t} - \lambda_{00}t}\right]
\end{multline}
\end{widetext}
we notice that there are three time integrations that result in delta functions,
but 4 $t$-like variables. In the case of the first term in the square brackets,
since it only appears in the combination $(\tilde{t} - t)$, this eliminates a variable,
and there are sufficient constraints to make it only depend on $\omega$ variables. However,
in the new term coming from flavor-mixing of the b-SYK, this is not true any more.
This is a signal of a breakdown of the ansatz Eq.~\eqref{eq:Lyapunov_Ansatz}.
We thus see that for consistency we must impose that $\lambda_{00} = \lambda_{10}$.
By repeating the argument for the other components of $\mathcal{F}$,
it can be shown that all Lyapunov components should be the same, $\lambda_{ij} = \lambda$,
and that there is only one Lyapunov exponent governing the behavior of the model.

\section{Recovery of the maximal Lyapunov exponent of the regular SYK}
At $\kappa=1$, the numerics reflect that the Lyapunov exponent of the model is the same as the maximal value of regular SYK. This can be understood by looking at the kernel Eq.~\eqref{eq:kernel_tau}. At $\kappa=1$, the scaling dimensions of both the $A$ and $B$ majoranas become $\frac{1}{4}$, and hence $G^A(\tau) = G^B(\tau) \equiv G(\tau)$, the 2 point function of regular SYK. The kernel then factorizes into the product of a function of the four imaginary times, and a constant matrix. 
\begin{equation}
    K(\tau_1\cdots\tau_4) = -J^2 G(\tau_{13})G(\tau_{24})G(\tau_{34})^2 \begin{pmatrix}
        1 & 2 \\
        2 & 1
    \end{pmatrix}
\end{equation}
The constant matrix in question has eigenvalues $-1$ and $+3$. The latter eigenvalue makes the kernel mathematically the same as the one for regular SYK, and hence the Lyapunov exponent should be the same.  Furthermore, it is for this reason that the special case of $\kappa=1$ allows the kernel to be diagonalized in the basis of the conformal blocks labeled by $h$. For $\kappa \neq 1$, the four components of the kernel transform differently under transformations of the conformal group.

\begin{figure}[h]
    \centering
    \includegraphics[width=1.0\columnwidth]{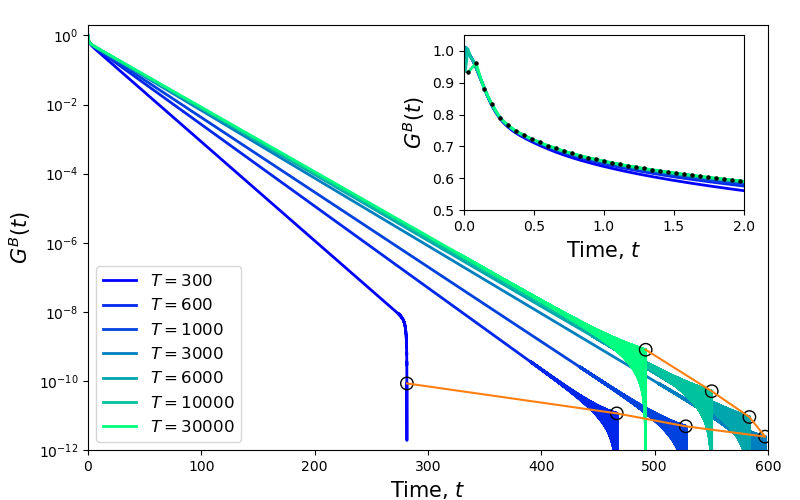}
    \caption{The Green function $G^B(t)$ for $\kappa=0.3$, $\beta=10$, $J=10$. 
      The number of discretization points is fixed to $N=2^{19}$ and the length of the time-grid $T$ is varied.
      The figure shows that increasing $T$ gives a better estimate of the decay-time for the Green function, but if $T$ is taken to be too high, numerical accuracy of the Green function is lost. The sweet-spot is here at $T=3000$.
      Black circles marks the position of (the point before) the first negative $G$, and is an estimate of the size of the numerical noise.     
    }
    \label{fig:FiniteSize}
\end{figure}

\section{Finite-size dependence of two-point functions and Lyapunov exponents}\label{app:FiniteSize}
In this section, we briefly comment on the sensitivity of the two-point function to the finite-size cut-offs introduced when numerically solving the Schwinger-Dyson equations for the b-SYK model.
To solve the coupled b-SYK equations (Eq.~\eqref{eq:retarded_dyson_equation} and below), we discretize the semi-infinite positive timeline by introducing a long time cut-off $T$ and a finite number of time steps $N$ inbetween. 
This introduces a discretized time-step $\Delta t=T/N$ and frequency step $\Delta \omega = 2 \pi / T$.
To avoid the discontinuities at $\omega=0$ and $t = 0$, we choose a time grid that is $t_n = \Delta t \cdot (n+1/2)$, and similarly for the frequency grid.

We can study of the effects of varying $T$ and $N$ on $G^B(t)$.

In Figure~\ref{fig:FiniteSize} we show an example for $\kappa=0.3$, $\beta=10$, and $J=10$.
We fix the number of discretization points to $N=2^{19}$ and plot $G^{B}(t)$ for several values of $T$. We have cut off the plot at the first negative value of $G^B$.
In the plot, we observe two qualitative effects of changing $T$: First, upon increasing $T$,
we find that the decay time (slope) of the Green function increases (decreases). Thus, increasing $T$, we allow $G^B(t)$ to behave as if the time axis was really semi-infinite.
One can perform a $1/T$ analysis and finds that the lines have a well-defined slope in the $T\to\infty$ limit.

Secondly, which is more subtle, we see that making $T$ too large decreases the quality of the approximation for $G^B(t)$, with the optimal number being around $T=3000$.
We arrive at this number by the following argument: In the plot, we only show $G^B(t)$ until the first non-negative value (at time $t_C$).
The solid-looking wedge shape that appears just before the first negative number is the effect of numerical oscillations that (as $G$ decreases) become relatively more important.
From the height where the ``wedges'' disappear (black circles connected with an orange line), we can approximate the size of this numerical error.
By inspection, we see that the smallest numerical errors (and also the largest $t_C$) happen for $T=3000$.
We can understand the loss by noting that as $T$ grows, then (for fixed $N$) $\Delta t$ also grows.
In the inset of the figure, one can see that at $T=30000$, $\Delta t$ is so large that it even affects the continuity of the curve $G^B(t)$.

Choosing the appropriate $T$, is thus affected by the range of the Green function decay, which in turn is affected by $\kappa$, the ratio between the two species.
In the numerics that we present in the main text, we worked with a fixed $N$ and $T$, which are good when $\kappa\approx1$ but not when $\kappa$ is increasingly asymmetric.
Errors in the two-point function will propagate and influence the calculations of the Lyapunov exponent and explain why we see the downturn of $\lambda$ at a characteristic value of $\kappa$.

\bibliography{bibSYK}

\end{document}